\documentstyle[12pt]{article}
\begin{document}

\begin{center}
{\bf Arguments on the Light-mass Scalar Mesons and Concluding Remarks\\
of the Meson Sessions}\\
S.F.~Tuan\\
{\em Department of Physics\\ University of Hawaii at Manoa\\ Honolulu,
HI 96822-2219 U.S.A.}\\

\vskip 1.0truein

\end{center}
\vskip 0.2truein
\begin{quotation}
This report attempts to summarize the most interesting (and hopefully
important) results leading up to and including those presented at the recent
Symposium sponsored jointly by the Institute of Quantum Science at Nihon
University and KEK. My task is to present the arguments on light-mass scalar
mesons below 1 GeV from both theory and phenomenological viewpoints,
including the new insight gained on $\pi-\pi$ production and scattering
amplitudes. Specific topics are taken up, particularly on the existence of
a $\sigma$(500-600) as explanation of the twin peak anomaly in $\Upsilon(3S)
\rightarrow \Upsilon(1S)\pi\pi$, the status of $J^{PC}=1^{-+}$ states and
a possible crypto-exotic hybrid with $J^{PC}=0^{-+}$ are discussed, as well
as the intriguing enhancement in $p\bar{p}$ radiative decay from $J/\psi$.
\end{quotation}

\begin{section}{\bf Introduction}
I wish to thank Professors Shin Ishida and Kunio Takamatsu san for assigning
me on July 25 last to talk on ``Arguments on the light-mass scalar mesons and
concluding remarks of the symposium'' at this great Symposium, thus postponing
one's UH retirement (except `on paper') in physics by some 7 months. It is an
opportunity to express my felicitations to Professor Ishida san upon entering
the age of GU-XI which C.N. Yang described as rare since ancient times, and
here is best wishes for the next 30 years! Not being an expert myself on
scalar mesons, I used the newsgroup approach to prepare myself for this talk,
garnering the expertise of worldwide physicists via extensive e-mails to help
me. I wish to express my deep appreciation to them all for their contributions
unstintingly given during the 7 months leading up to the Symposium. In the end
I had to follow the L.B. Okun principle of truth, namely talk only about those
subjects I personally am satisfied with. However I may be forgiven for errors
of commission or omission because after all T.D. Lee once said that 50\% of
theoretical physics (and I am a theorist) is emotion, and I am exercising this
privilege! In what follows I shall first discuss 5 theoretical options for
understanding light-mass scalar mesons below 1 GeV. This is followed by 
highlighting the new insights gained on $\pi-\pi$ production and scattering
amplitudes. Specific and selective topics as delineated in the abstract above,
are discussed mainly based on my familiarity with the subject matter. 
Concerning the status of $\sigma$ and $\kappa$ and their existence, Session V at
this Symposium was truly excellent. Ms. Carla Goble did an outstanding job at
presenting the experimental status of these states, and Muneyuki Ishida san
did a persuasive job on phase motion for $\sigma$ and $\kappa$ production
amplitudes. I will recommend everyone to read the Proceedings articles for
Session V [Properties and Spectra of Scalar Mesons I], since all need to
savor the contributions here, without the less than adequate explanation of
yours truly in summary. I conclude on the basis of Session V that both
$\sigma$ and $\kappa$ indeed exist.

\end{section}

\begin{section}{Theoretical Models}
We examine in this section 5 theoretical models which address the issue of
low mass scalars below 1 GeV, assuming that $\sigma$(500) and/or $\kappa$(800)
indeed exist. The options are:-

\begin{enumerate}
\item \underline{The [$\sigma,\kappa,a_{0}(980),f_{0}(980)$] form an usual
($q\bar{q}$) L=1 scalar nonet below 1 GeV.} This seems quite reasonable at first
sight, since it builds on the well known ($q\bar{q}$) L=0 nonet [$\pi,\rho,$
etc.] below 1 GeV also. There are however a number of \underline{problems}
with this approach as enumerated by Schechter\cite{REF1} sometimes back. They
are:-
\begin{enumerate}
\item
[$\bullet$] The $a_{0}$(980) and $\sigma$(500) have same number of non-strange quarks
        but curiously are NOT degenerate.
\item
[$\bullet$] As scalar P-wave states, why are they not in $> 1$ GeV energy region of
        other P-wave states.
\item
[$\bullet$] There is no explanation on why $f_{0}$(980) $\sim s\bar{s}$, and
        $a_{0}^{+}(980) \sim u\bar{d}$ are degenerate. Achasov\cite{REF2}
        regards explanation of $\phi \rightarrow f_{0}(s\bar{s})\gamma$ detached
        from $a_{0}$(980) as `awful' from a theoretical viewpoint.
\item
[$\bullet$] We do not deny that Scadron\cite{REF3} produced in the $q\bar{q}$
        approach, not only the famous Nambu relation $m_{\sigma} \simeq
        2m_{q}$, but also the treasured Sakurai vector meson dominance 
        universality condition. However $q\bar{q}$ (L=0) for low lying vector mesons
        has not been challenged. They lie in the domain of isolated 
        environment, remote from background, where QCD sum rules agree very well in
        the words of Achasov\cite{REF4}. The linear $\sigma$-model (Scadron),
        effective lagrangians, obtained in the Nambu-Jona Lasinio NJL type
        model are valid, strictly speaking, only for `virtualities' much less
        than the threshold for creation of $q\bar{q}$ pair, i.e. light (better
        still for massless) particles. Sometimes it works, e.g. in the vector
        channel\cite{REF4}. I am inclined to agree with Bjorken\cite{REF5} on
        $m_{\sigma} = 2m_{q}$ that ``Sorry, but I don't have anything useful to
        say about $\sigma$ mass. It is pushing the limits of what NJL can do
        in my humble opinion''.
\item
[$\bullet$] One notes that Van Beveren, Scadron, {\em et al.}\cite{REF6} argued
        strongly against the $q\bar{q}$ approach of Shakin-Wang, thus further
     attenuating a common $q\bar{q}$ modeling of low mass scalars.
\end{enumerate}
\item \underline{What if $\sigma$ exists $\leq$ 1 GeV, but $\kappa$(800) does not
exist\cite{REF7}.} In Narison's words\cite{REF8} ``In addition to the glueball
of mass 1.5 GeV found e.g. on the lattice in the quenched approximation as well
as from sum rule, a light glueball is needed for simultaneous saturation of the
subtracted and unsubtracted sum rules where the stablization scale occurs at
different regions due to the important role of the subtraction constant of the
two-point correlator in the subtracted sum rule. Contrary to the one at 1.5
GeV which has a U(1) glueball like decay ($\eta-\eta^{\prime}$,...), this
{\em light} glueball couples strongly to $\pi-\pi$ (OZI violation) and is not
found from quenched lattice simulation. A more accurate estimate of this mass
on the lattice would need the inclusion of quark loops''. I believe his recent
e-print\cite{REF8} discusses this point further. Ochs and Minkowski\cite{REF9}
appear to endorse this viewpoint, since $f_{0}$(980)/$a_{0}$(980) can be
returned to the $q\bar{q}$ assignment together with above 1 GeV scalars
$f_{0}$(1500) and $\kappa$(1430). Of course the case here will be greatly
weakened if $\kappa$(800) is experimentally confirmed.

\item \underline{$(qq)(\bar{q}\bar{q})$ - states.} This structure allows
two configurations in color space: $\bar{3}3$ and {$6\bar{6}$}. They may mix/rearrange to form
($q\bar{q})(q\bar{q})$ with color configuration, and there could be difficulty to
distinguish a tetraquark state from a mesonic molecule. Isgur and Weinstein
observed that four quarks in a confining potential are mainly arranged as two
color singlets at large (1.5 fm) distance and give birth to \underline{MESONIC
MOLECULES.\cite{REF10}} Such a loose bound structure appears consistent with
Rosner's statement\cite{REF11}:- ``A low-mass $(\sim$ 600 MeV) scalar (as a
dynamically induced $\pi-\pi$ interaction) is consistent both with theory
(current algebra, crossing symmetry, and unitarity) and with experiment. A
similar enhancement appears to be occurring in the I=$\frac{1}{2}$ S-wave
$K-\pi$ system. Indeed it may even be relevant to the electroweak scale if
the Higgs boson turns out to be a broad, dynamically generated object rather
than the narrow state everyone is hoping for at 115+ GeV.'' Adler\cite{REF12}
is supportive of Rosner, namely the {\em PCAC} consistency condition implies
the need of a low energy and broad $\pi-\pi$ scattering resonance.

\underline{CAVEAT!} Theorists have done sophisticated work in building 
analyticity, unitarity, and crossing symmetry into dispersion theoretic and Roy
Equations procedures\cite{REF13} either \underline{for} the existence of
$\sigma$ or \underline{against} its existence. A note of caution has been
introduced by Anisovich\cite{REF14} concerning the \underline{left-hand cut
problem}, to wit:- ``.... The $\pi-\pi$ amplitude has a left-hand cut affected
by interaction forces, so constraints coming from left-hand cut are important.
Our knowledge is however rather vague (mostly to test \underline{individual}
hypothesis e.g. $\rho$-exchange etc.). The left-hand cut is in fact highly
unstable near $\pi-\pi$ threshold s. A comparatively small change of partial
wave amplitudes affect the left-hand cut considerably. The reason being the
contributions from different (t and u) channels and different resonances
\underline{cancel each other out to a great extent.}'' Thus fitting data
within rigid constraints on $\pi\pi$ amplitudes for s $< 4m_{\pi}^{2}$ can lead
to incorrect amplitude representations in the mass region of $\sigma$ searched
for! In my humble opinion theory methods here may not be definitive concerning
the existence/non existence of $\sigma$-meson.

\item \underline{Ishida Model of Chiral Particles/Covariant Level Classification.}
T\"{o}rnqvist\cite{REF15} in his Kyoto summary commented on Professor Ishida
san's relativistic S-wave $q\bar{q}$ bound states (``chiralons'') with $J^{PC} =
0^{++}$ quantum numbers. They would appear only at low masses, thus the
possible low mass scalars ($\sigma,\kappa,a_0{(980)},f_0(980)$) below 1 GeV could
be potential candidates for the chiralon model. He allowed that the model is
very speculative, but would open up a new approach. Since the low mass scalars
enumerated here can be given alternative explanations, I asked \underline{where
is the `smoking gun' test of the Ishida model?} and concentrated on the 
prediction of a scalar {\em $B_{0}^{\chi}$} in the B-meson system, because it has
been proposed\cite{REF16} as possibly relevant to the explanation of the double
peak anomaly in $\Upsilon(3S) \rightarrow \Upsilon(1S)\pi\pi$ also. Ito
{\em et al.}\cite{REF17} pointed out that the {\em $B_{0}^{\chi}$} has a mass of
$\sim 5550$ MeV, hence ${\em B_{0}^{\chi}} \rightarrow B+\pi$ (S-wave) is
allowed, as they analysed from the data of L3 and Aleph. Also at this 
Symposium Yamauchi\cite{REF18} suggested a width $\sim 20$ MeV for {\em $B_{0}^{\chi}$}
state. Clearly it is important to have independent affirmation of this state
by an experimental group, e.g. at CLEO where much \underline{fine} features
of B-physics are done. I did push my CLEO friend Sheldon Stone very hard on
this since July, 2002. However Sheldon's reply\cite{REF19} is ``I don't have
much to add. The new CLEO $\Upsilon$(3S) data shows the same double peak
structure as before. I don't think we are going to be able to say much about
{\em $B_{0}^{\chi}$}. Have fun in Tokyo. Sheldon''.

\item  \underline{Mesonic Molecules Revisited.}
Inspired by Isgur-Weinstein\cite{REF10} $(qq)(\bar{q}\bar{q})$ case but considering
only mesonic degree of freedom (i.e. color singlets), e.g. $\rho$-exchange
between $K\bar{K} \leftrightarrow f_{0}(980)$, a very compact molecule can be
obtained. The reason is that radiative $\phi$ decay data exclude the extended
loosely bound Isgur-Weinstein molecule, while giving strong evidence for a
compact $K\bar{K}$-state or a \underline{compact four-quark state}\cite{REF19}.
I think the Adler-Jaffe-Achasov approach\cite{REF12} could be a mix of option
3. above and option 5. here. We need a broad $\sigma$: $\pi-\pi$ resonance at
low mass \underline{and} would like to accomodate $f_{0}(980)$ as a compact
four-quark state - \underline{to form below 1 GeV scalar nonet.}
\end{enumerate}
Anisovich argues that classification of Kaon states on a (J, $M^{2})$-plane
\cite{REF20}, points out that a $\kappa (\sim$ 900 MeV), much under discussion
at this Symposium, {\it does not belong to (J, $M^{2})$ trajectories
related to $q\bar{q}$ states. This could be a strong argument against $\kappa$ as
$q\bar{q}$ since all other $q\bar{q}$ states lie on these linear trajectories well.}
We can turn this argument around. A $\kappa$(800-900 MeV) cannot be a glueball
(isospin); it is implausible to be a hybrid (believed to be significantly
higher than 1 GeV in mass). What else can it be if existence is established?
A {\em reductio ad absurdum} mathematical reasoning would conclude:- {\bf
$\kappa$ is a multi-quark state!}

\underline{Important Point}: The quark content of ($\sigma,f_{0}(980),
a_{0}(980),\kappa$) are ``virtuality'' dependent in the Bogoliubov sense.
Adler\cite{REF21} calls it coherent states which is worth following up in the
future. The developments at Hadron 2001\cite{REF12} have persuaded
T\"{o}rnqvist\cite{REF22} to be no longer troubled that the pion could end
up as a 4q state. He has embraced a viewpoint proposed earlier by Kunihiro and
Nambu\cite{REF23} that these low lying scalars are \underline{emerging}
Goldstone Higgs nonet (of strong nonperturbative interactions when a hidden
local symmetry is spontaneously broken) and is a superposition of $q\bar{q},
qq\bar{q}\bar{q}, qqq\bar{q}\bar{q}\bar{q}$,... The word `emerging' was coined from a
Bjorken paper\cite{REF24} where Jaffe's natural nonlinear realization relation\cite{REF12}
\begin{eqnarray}
\sigma = [f^2-\vec{\pi}^2]^{1/2} = f - \vec{\pi}^2/2f +.....
\end{eqnarray}
\noindent is reproduced in Bjorken's Eq. (12)\cite{REF24}. Thus there is a
convergence of views!

\underline{Observations:}
\begin{enumerate}
\item[$\bullet$] If the scalar nonet is predominantly of the Jaffe 4q-state $(3,\bar{3})$
        coupled in color to form a picture of a ``cryptoexotic'' nonet below
        1 GeV, than an inverted equally spaced spectrum follows\cite{REF22}.
        The currently touted low energy scalars $a_{0}(980)/f_{0}(980)$,
        $\kappa(\sim 800)$, $\sigma(\sim 600)$ follow this spacing rule
        remarkably well. The physical picture is that of a four quark 
        component in the core transforming to a meson-meson description in the
        periphery as first proposed by Jaffe and Low\cite{REF25}.
\item
[$\bullet$] How about instanton effects that Narison\cite{REF8} emphasizes as
        important at least in the context of the $\eta^{\prime}$ problem. Could
        it cause trouble for the multi-quark approach as proposed for instance
        by Adler-Jaffe-Achasov\cite{REF12}? Kochelev\cite{REF26} maintains that
        ``From his point of view, a multi-quark interaction, induced by
        instantons, could be the \underline{reason} for the observed 
        enhancements in $0^{++}$ channels.''
\item
[$\bullet$] Iwasaki\cite{REF27} in connection with `Meson Bound-State Experiment'
        in a nuclear medium says that $\sigma$ below 280 MeV (two pion
        threshold) would become stable (hence very narrow width). This seems to
        resonate with the work of Jaikumar and Zahed\cite{REF28} which
        discusses scalar-isoscalar excitation in the different context of
        dense quark matter (up to neutron star density). Here $\sigma$-meson
        in this (color-flavor locked) phase appears as a 4q state (diquark
        and anti-diquark) with a well-defined mass and extremely small width,
        as a consequence of its small coupling to two pions. My question is
        that in the A-J-A inspired paper\cite{REF12}, it was argued that when
        Bogoliubov virtuality of $\sigma$-state is $\sim m_{\pi}$ mass, the
        state has $q\bar{q}$ characteristics, as the chiral partner of $\pi$. But
        when virtuality of $\sigma \leq$ 1 GeV, it exhibits 4q state 
	characteristics. Is Jaikumar and Zahed\cite{REF28} saying that even below the
        2$\pi$ threshold (significantly closer to $m_{\pi}$ than to 1 GeV),
        $\sigma$ continues to retain 4q characteristic? Perhaps the 
        experimental work of the Dirac Collaboration\cite{REF29} on Pionium, Kaonium
        will clarify this issue.
\end{enumerate}
\begin{center}{\underline{The Future}}
\end{center}

Beyond the light scalars $< 1 GeV$, lies the $U(3) \times U(3)$ linear $\sigma$
model (L$\sigma$M) where T\"{o}rnqvist\cite{REF15} identified Joe Schechter
as one of the originators. This model\cite{REF30} does not exist in the tree
approximation (unfortunately!). The model is renormalizable and strategically
(my favorite word) its study on the lattice above 1 GeV say, is very intri-
guing. {\bf But the existence of $\kappa(800)$ is of course critical to the
theory here under our option 5.}

%\end{section}

\section{Significant New Insight}
Muneyuki Ishida san\cite{REF31} makes an \underline{important} critique of
conventional analysis concerning $\pi\pi$ production amplitude $\cal F$ and
scattering amplitude $\cal T$ usually written as
\begin{eqnarray}
{\cal F}_{\pi\pi} = \alpha (s) {\cal T}_{\pi\pi}; \alpha (s): \ \mbox{slowly varying real
function}.
\end{eqnarray}
\noindent This implies that $\cal F$ and $\cal T$ have the same phases and the
same structure. There is also the implication of \underline{common positions
of poles,} if they exist.

It has often been argued that the amplitudes for $J/\psi \rightarrow
\omega\pi\pi$ and  $D^{+} \rightarrow \pi^{-}\pi^{+}\pi^{+}$ (for $\pi\pi$ 
production) \underline{must assume same phase} as the $\pi\pi$ scattering phase,
since to take the case for $J/\psi$ the energy range $m_{\omega\pi} (\sim
M_{J/\psi})$ is large, and $\pi\pi$ decouples from $\omega$ in the final state
channel. Hence phase constraint is argued to come from $\pi\pi$ elastic
unitarity condition [Fermi-Watson-Migdal Theorem].

However according to the excellent work of Suzuki-Achasov\cite{REF32}, large
relative strong phase is needed for $J/\psi \rightarrow VP (\omega\pi^{0},
\omega\eta^{0},\rho\pi,K^{*}\bar{K},$ etc.) for the famous puzzle here, likewise
for $J/\psi \rightarrow PP.$ We are in the domain of long distance effect and
\underline{non} perturbative QCD regime. Suzuki\cite{REF33} put it succinctly
as `In $J/\psi \rightarrow \omega\pi\pi$, the invariant $m_{\omega\pi} \leq
1.6 \ GeV.$ There are ``overlapping resonant phases'', in fact many non strange
resonances up to 2 GeV in the PDG Table can contribute. For $D \rightarrow
K\pi$, the $K$ and $\pi$ are back to back at 1.87 GeV. The CLEO experiment
says\cite{REF34} that $K-\pi$ interaction is very large $(\sim 90^{o})$ in final
state interaction phase difference $\delta_{\frac{3}{2}}-\delta_{\frac{1}{2}}$
at $m_{D}$. Hence $K-\pi$ scattering at 1.87 GeV is definitely \underline{not}
in perturbative QCD regime. The same is true for $K^{*}-\pi$ at 1.87 GeV.'
The situation in $B \rightarrow D\pi,D\rho,D^{*}\pi$ has been discussed
recently by Rosner-Chiang\cite{REF35}.

If $\cal F$ = $\alpha(s)\cal T$ is not true for say $J/\psi \rightarrow
\omega\pi\pi,\pi K\pi K$ and $D \rightarrow \pi\pi\pi,K \pi\pi$, even though the
question of rescattering and its importance remains debatable, we need to
\underline{separate} analysis involving (n $\geq$ 3) multi-hadron final states as
listed here from the clean n=2 case. To wit, following Ishida\cite{REF31},
let us take the FOCUS experiment\cite{REF36} $D^{+} \rightarrow
K^{-}\pi^{+}\mu^{+}\nu$ with n=2. The $K^{-}\pi^{+}$ piece is isolated at strong
interaction level, hence Fermi-Watson-Migdal FWM theorem is applicable. This
in turn implies that $\cal F$ and $\cal T$ have common phase.
FOCUS\cite{REF36} says that $D^{+} \rightarrow \bar {K}^{*0}\mu^{+}\nu$ is
dominant, with a small S-wave $(K^{-}\pi^{+}$ component, which exhibits a more
or less \underline{constant phase $\delta=\frac{\pi}{4}$ in the region of
$\kappa$ with $m_{K\pi} \sim 0.8 \ GeV.)$} This $\delta$ is suggested by Ochs
and Minkowski\cite{REF9} to be the same as $K-\pi$ scattering phase shift by
the extensive LASS data\cite{REF37}. Hence FWM theorem is respected. It is of
interest to note that Shabalin\cite{REF38} studied $K_{e4}$ decay, using SU(3)
and L$\sigma$M, by treating $K^{+} \rightarrow \pi^{+}\pi^{-}e^{+}\nu$ as
$K^{+} \rightarrow \sigma e^{+}\nu (\sigma \rightarrow \pi^{+}\pi^{-})$
to explain the large width of $K_{e4}$ decay (twice as large as soft
pion prediction). Here also the amplitude observes the FWM theorem and has the same phase
as $\pi\pi$ scattering phase shift; the $\sigma$ Breit-Wigner phase motion is
not observed, but the large width suggests nevertheless that $\sigma$ is being
produced! This case is quite analogous to the FOCUS situation discussed above.

Take now the E791 experiment\cite{REF39} (with n=3) on $D^{+} \rightarrow
(K^{-}\pi^{+})\pi^{+}$. Here $K^{-}\pi^{+}$ piece is not isolated in strong
interactions (rescattering or other mechanism can prevail) and the FWM theorem
becomes questionable. Indeed E791 sees large phase motion and Breit-Wigner in
the $\kappa$(800) region. Evidently production amplitude ${\cal F}_{K\pi}$ can
have different phase from $\cal T_{K\pi}$ of scattering. I am impressed that
Ms. Carla Goble, who participated in both E791 and FOCUS experiments, reported
at this Symposium that there is no contradiction between E791 and FOCUS.

Though Carla has no doubt improved the sophistication of analysis since the
E791 publication\cite{REF39}, I am attracted by Brian Meadows\cite{REF40} and
his statement that `E791 did an excellent job in showing that a ``NR''
(contact) amplitude with magnitude and phase independent of position on the
Dalitz plot simply did not work as it had for earlier data. The next simplest
assumption, \underline{a S-wave Breit Wigner,} did however give an excellent
description of the data, \underline{both magnitude and phase} - in both $K\pi$
and $\pi\pi$ systems. Really that is all.' I applaud this as a most sensible
way to proceed - \underline{the next simplest assumption.} I urge
CLEO\cite{REF41} which studies $D^{0} \rightarrow K_{S}^{0}\pi^{+}\pi^{-}$ with
a currently \underline{null} result on $\kappa^{-}(800) \rightarrow
K_{S}^{0}\pi^{-}$, and BABAR\cite{REF42} with a \underline{null} result on
$\kappa^{+}(800)K^{-}$ in $D^{0} \rightarrow K^{0}K^{-}\pi^{+}$ Dalitz plot
analysis, to follow the next simplest assumption to verify/refute E791. The
Breit-Wigner and phase motion should be similar to E791 in CLEO/Babar 
according to Ishida\cite{REF31}.

\underline{Remark:} Bugg\cite{REF43} maintains that `Particles are to be
identified as poles. If you allow the pole position to be different in 
different processes, you are allowing the particles to have different mass and
width in different processes. This is \underline{not} true for $\rho$(770) and
$f_{2}$(1270) in different processes......' However $\rho$(770) and
$f_{2}$(1270) could again be a case of isolated environment, remote from 
background\cite{REF4}. Indeed our experience\cite{REF11} with reactions
$\gamma\gamma \rightarrow \pi\pi$ indicates that the way in which a
\underline{broad} resonance (like $\sigma$, $\kappa$) \underline{manifests}
itself can differ significantly from process to process, particularly compar-
ing elastic versus inelastic channels. I think Eric Swanson's comment at the
end of my Symposium talk best summarizes this point. Namely, `the position of
a pole in the S-matrix is a fixed property of the underlying field theory. Thus
its position cannot depend on the observable used to obtain it (of course, one
must extrapolate to obtain pole positions, so numerical+experimental error may
lead to different numbers). On the other hand, the Breit-Wigner formula is an
approximation to what may be a complicated experimental situation and BW
parameters (such as resonance mass and width) can change depending on the
channel. However, one expects these variations to be small if the resonance is
narrow.' From the compilation of $\sigma$, $\kappa$ poles\cite{REF44} they
would appear to be significantly broader than $\rho$(770) and $f_{2}$(1270).

If $\sigma$ exists at low mass, the 50\% emotional component of the theorist
in me, would urge that it be deployed to solve other outstanding puzzles. Here
the phenomenological work of the Ishida group\cite{REF45} \underline{excites}
me. Remember multipole expansion of QCD has more or less successfully treated
the suppression of spectra near $\pi\pi$ threshold [the resulting amplitude
has Adler zero around $s \sim 0$] in
\begin{eqnarray}
\psi(2S) \rightarrow \psi(1S)\pi\pi  \ \Delta E & = & 589 MeV   \ ka
\sim 0.7  \nonumber \\
\Upsilon(2S) \rightarrow \Upsilon(1S)\pi\pi  \ \Delta E  & = & 563 MeV
\ ka \sim 0.3 \nonumber \\
\Upsilon(3S) \rightarrow \Upsilon(2S)\pi\pi  \ \Delta E & = & 332 MeV
\ ka \sim 0.18
\end{eqnarray}
\noindent here k is typical momentum of the emitted gluons and $a$ is size of
$Q\bar{Q}$ system estimated from potential models, with k given by\cite{REF46}
\begin{eqnarray}
k \sim \frac{1}{2}(m_{\Phi^{\prime}} - m_{\Phi}), \Phi = \psi, \Upsilon
\end{eqnarray}
for two gluon emission. Similar results are obtained if we choose
$\frac{1}{k} \sim 1 fm$, as typical size of light hadrons. It is also known
from nuclear physics\cite{REF46} that the classification of multipole orders
is valid even for $ka \sim 1$. The \underline{fly in the ointment} is in
\begin{eqnarray}
\Upsilon(3S) \rightarrow \Upsilon(1S)\pi\pi  \ \Delta E = 895 MeV  \ ka \sim 0.48
\end{eqnarray}
\noindent where the $\pi\pi$ spectrum exhibit the double peak anomaly. Why is
the multipole approach failing for such reasonable $ka \sim 0.48$ value? The
Ishida group\cite{REF45} proposes the tantalizing intervention of the $\sigma$
in this dipion mass range $0 < m_{\pi\pi} \leq 900 MeV$. Actually they 
introduce the production amplitude ${\cal F}_{2\pi}^{G}$ (where G depict 
ground-state tensor and scalar intermediary glueballs, which seem a straightforward
extension of two gluon intermediary of multipole/QCD work) and take form
\begin{eqnarray}
{\cal F}_{2\pi}^{G} \approx -2\xi^{G}p_{10}p_{20}, \ \mbox{vanishes when} \ p_{1\mu}
\rightarrow 0_{\mu}.
\end{eqnarray}
\noindent Thus, it has Adler zero, satisfies general constraint from chiral
symmetry, and does not vanish at $s = m_{\pi}^{2}.$ In a relevant process,
Adler limit $p_{1\mu} \rightarrow 0_{1\mu}$ implies neglect of $\Delta E$ in
comparison with $m_{\pi}$. For $\Upsilon(3S) \rightarrow \Upsilon(1S)\pi\pi$
transition at dipion threshold $s = 4m_{\pi}^2$, $p_{1\mu} \cong p_{2\mu}$,
which in turn implies $p_{10} = p_{20} \approx \frac{\Delta E}{2} = 450 MeV
\gg m_{\pi}$. Hence ${\cal F}_{2\pi}^{G}$ is almost s-independent in the physical
region, and has no zero close to threshold. There is therefore no suppression
near threshold, and the $\pi\pi$ spectrum can show steep increase from its
threshold. Including both ${\cal F}_{\sigma}$ Breit-Wigner for $\sigma$ and a
${\cal F}_{2\pi}$ for a direct $2\pi$ amplitude, we can write the phenomenology
as an amplitude
\begin{eqnarray}
{\cal F}^{phen} \equiv {\cal F}_{\sigma+2\pi}^{phen} + {\cal F}_{2\pi}^{G}.
\end{eqnarray}
\noindent \underline{Remarks:}
\begin{enumerate}
\item
[$\bullet$]Constraints of chiral symmetry require that the ${\cal F}_\sigma$ 
amplitude must be strongly cancelled by non-resonant \underline{repulsive}
$\pi-\pi$ amplitude in $\pi\pi$ scattering, i.e. destructive interference
between ${\cal F}_{\sigma}$ and ${\cal F}_{2\pi}$ explains suppression of threshold
spectra for $\Upsilon(3S) \rightarrow \Upsilon(2S)\pi\pi$, $\Upsilon(2S)
\rightarrow \Upsilon(1S)\pi\pi$, and $\psi(2S) \rightarrow \psi(1S)\pi\pi$.
For $\Upsilon(3S) \rightarrow \Upsilon(1S)\pi\pi,$ these amplitudes interfere
\underline{constructively}. These behaviors of production amplitudes are 
consistent with chiral symmetry constraint. Of special interest is that the valley
between double peak $\pi\pi$ spectrum structure in $\Upsilon(3S) \rightarrow
\Upsilon(1S)\pi\pi$ appears correlated to where $\sigma$ reaches peak position.
Hence the spectrum is well reproduced by interference between direct
${\cal F}_{2\pi}$ with zero phase and ${\cal F}_{\sigma}$ with moving phase.
{\it We are in fact observing the very phase motion of 
$\sigma$-Breit-Wigner via this double peak anomaly puzzle!}
\item
[$\bullet$] Bugg\cite{REF43} understood Ishida group's background phase shift
$\delta_{BG}$ to be \underline{repulsive} in the following way. Take for
orientation $\sigma: M-i\frac{\Gamma}{2} \sim 550-i250 MeV$. This implies a
scattering length $\sim \frac{1}{m_{\pi}}.$ But the $K_{e4}$ data demand a
scattering length $\sim \frac{0.22}{m_{\pi}}.$ Hence repulsive short range
interaction is needed to balance $\sigma$ resonance contribution. Here at the
Symposium M. Ishida san answered the mechanism for such a repulsion as due to
model independent chiral cancellation.
\item
[$\bullet$] Ishida san\cite{REF47} suggested that the ${\cal F}_{2\pi}^{G}$ 
derivative type interaction may have other origin. For instance all sequential two-
pion production gives this type of amplitude according to the Adler self-
consistency condition. Indeed the explanation of double peak anomaly by 
postulating the existence of $X(b\bar{bq}\bar{q})$ resonance with S-wave decay into
$\Upsilon(1S)-\pi$ in the 10.4-10.8 GeV mass range\cite{REF48} does include
the sequential two-pion production diagram. Hence in a partial sense this
approach is consistent with the Ishida group\cite{REF45} analysis of the double
peak anomaly in $\Upsilon(3S) \rightarrow \Upsilon(1S)\pi\pi$. The convergence
to seek an unified viewpoint (and hence an unique interpretation) of the
anomaly is far from complete however, since Anisovich {\em et al.}\cite{REF48}
count heavily on the importance of the rescattering triangle graph (leading to
logarithmic singularity) in their work. Indeed in a recent communication 
Anisovich\cite{REF14} again maintains the possible existence of logarithmic
(triangle) singularities near $\pi\pi$ threshold, and they should be taken into
account in search of $\sigma$ meson in three particle process. The influence
of such singularities on $\pi\pi$ spectra are important for low $\pi\pi$ mass
(e.g. at $f_{0}$(980)) and surely will be more so at the lower $\sigma$ mass.

From an experimental point of view the $X(b\bar{bq}\bar{q})$ model\cite{REF48} is
difficult to establish. We need $\Upsilon(nS) \rightarrow \pi+\Upsilon(1S)+\pi$
with $n \geq 6$ to do an adequate $\pi + \Upsilon(1S)$ mass distribution plot.
On the other hand Sheldon Stone\cite{REF49} at CLEO has recently expressed
interest in the $\sigma$ explanation when he writes `If the $\sigma$ 
explanation is correct, are there any other physical variables predicted besides the
$\pi-\pi$ mass spectrum? Any angular distributions etc...?' I leave this as
a challenge for the Ishida group to work out.
\end{enumerate}
\end{section}

\section{Selected Meson Topics of Symposium Beyond 1 GeV}

Here I will take liberty by commenting only on those states above 1 GeV for
which I have some familiarity from the past. Hence my apologies to Symposium
speakers whose very good work I am unable to summarize because of my own lack
of experience in their subject matter.

The $J^{PC} = 1^{-+}$ state has long intrigued me. At Protvino (Hadron, 2001)
Ted Barnes\cite{REF50} in his summary talk concluded that the VES collaboration
(V. Dorofeev, Hadron 2001) had no clear preference for a $\pi_{1}(1400)$ (this
is the M(1405) of GAMS experiment in 1988 with $I=1, J^{PC}=1^{-+}$) resonance
interpretation. Fit of similar quality can be obtained from a non resonant
signal. Hence the critique of Dalitz {\em et al.}\cite{REF51} about P-wave
interfering background of GAMS(1988) still has some vitality. The
E852/Brookhaven (A. Popov, Hadron 2001 Proceedings) favors the C-exotic
$\pi_{1}(1600) \rightarrow \eta^{\prime}-\pi, \rho-\pi,$ and $b_{1}(1235)-\pi$.
Nevertheless for a hybrid $q\bar{qg}$ state, theory [Flux
tubes\cite{REF50}, 
confining field theory\cite{REF52}] continues to maintain a mass prediction for
$1^{-+}$ between 1.9-2.1 GeV! Given my comfort with multi-quark scalars below
1 GeV, I ask \underline{why not treat} $\pi_{1}(1600)$ as the lightest(?)
C-exotic from multi-quark (e.g. 4q) configuration? It is good to hear from
Tsuru san at this Symposium a historical review of the search for $J^{PC} =
1^{-+}$ states since 1988. It is also pleasing to hear from Director Alexander
Zaitsev\cite{REF53} that VES search in $\pi^{-}A \rightarrow
[b_{1}(1235)\pi]A$ supports a four quark interpretation for $\pi_{1}(1600)$ in
$\pi^{-}p \rightarrow \eta\pi{n}, \eta^{\prime}\pi{n}$ charge exchange. His
suggestion of the existence of $\pi(1800)$ with $J^{PC} = 0^{-+}$, where
$\pi(1800) \rightarrow f_{0}(980)\pi, \epsilon(\sigma)\pi, a_{0}\eta,
f_{0}(1500)\pi$ [mostly unitary singlet + octet?] but \underline{NOT} to
$\rho\pi$ [octet + octet] points towards a crypto-exotic hybrid $q\bar{qg}$
(longitudinal fluctuation of flux tube). This intriguing possibility surely
deserves further study.

Steve Olsen\cite{REF54} at this Symposium discussed the observation of a near
-threshold enhancement in the $p\bar{p}$ mass spectrum from radiative $J/\psi
\rightarrow \gamma{p} \bar{p}$ decays, using the 58 million $J/\psi$ sample at
BES. Fitted as an S-wave, the peak mass is below $2m_{p}$ (1876.54 MeV) at
1859 MeV with total width $\Gamma < 30 MeV$ (90\% C.L.). They are not yet seen
in $J/\psi \rightarrow (p\bar{p})\pi^{0}, (p\bar{p})\eta^{0}$ which in
Ishida language could mean the $p\bar{p}$ are not isolated at the strong interaction level
(n=3 final state) in contrast to $\gamma{p}\bar{p}$ where $p\bar{p}$ is isolated
in strong interaction (n=2). Note at BELLE\cite{REF55} $B \rightarrow
(p\bar{\Lambda)}\pi$ threshold enhancement is seen for $p\bar{\Lambda}$. Here the
$\pi$ from phase space consideration may have moved significantly away from
the $p\bar{\Lambda}$ environment. Implications of the Olsen discovery has been
thoroughly analysed by Rosner\cite{REF56}, hence I will comment only on some
items of particular interest to me.
\begin{enumerate}
\item
[$\bullet$] The threshold enhancement of $p\bar{p}$ in $J/\psi \rightarrow
\gamma{p}\bar{p}$ is not due to Coulomb attraction as Olsen showed at this
Symposium.
\item
[$\bullet$] The Nambu relation $m_{\sigma} = 2m_{q}$ mentioned earlier should
strictly be applied to the nucleon mass as $m_{\sigma} = 2m_{p}$, since the
Nambu-Jona Lasinio papers\cite{REF57} were written some two years earlier than
the invention of the quark model. If the threshold enhancement\cite{REF54} is
treated as a $J^{PC}=0^{++}$ scalar state, the fit yields a peak mass $2m_{p}$
and a very narrow total width $\sim$ 4.6 MeV.
\item
[$\bullet$] If the S-wave interpretation with $J^{PC} = 0^{-+}$ is correct,
Suzuki\cite{REF33} points out that they could be baryon-antibaryon six quark
molecular states $(qqq\bar{q}\bar{q}\bar{q})$ where the $(qqq)$ and
$(\bar{q}\bar{q}\bar{q})$ are pulled together by the gluon equivalent of
the molecular Van der Waals force (analogous to the suspected charmonium $D\bar{D}$
molecular state mode of $cc\bar{c}\bar{c}$). The emphasis is then on the deuteron
analogy, and not on Fermi/Yang or Sakata model dynamics.

%\end{section}
\end{enumerate}

\section{Conclusion}

We have seen that a credible picture for the existence of a nonet of scalars
$[\sigma,\kappa,f_{0}(980),a_{0}(980]$ has emerged from this Symposium.
Indeed perhaps a multi-quark interpretation of these states has once again
become an attractive possiblity for theoretical understanding. The natural
question to ask is `how about the baryons'. At the end of Professor Oka san's
Symposium summary on the baryon system, I asked whether in connection with
the $\Lambda(1405)$, it could also be a case of what Steve Adler\cite{REF21} calls co-
herent states $qqq, qqqq\bar{q},.....$ in the Bogoliubov sense as applied to
baryons. I believe Pakvasa and I actually alluded to this possibility in our
paper\cite{REF58} together with some other baryon examples, though the
clarification for mesons (as coherent states) only came later at
Protvino\cite{REF12}. We have certainly come a long way from the traditional
naive quark model classification of hadron states\cite{REF59} of some 37 years
ago. Hadron 2001 at Protvino and now at this Tokyo Symposium have jointly
contributed towards an exciting and challenging future for hadron physics!
In addition to the general acknowledgement given in the Introduction, this
work was supported also in part by the U.S. Department of Energy under Grant
DE-FG-03-94ER40833 at the University of Hawaii at Manoa.

%\end{section}


\begin{thebibliography}{99}
\bibitem{REF1} J.~Schechter, Proceedings of $\sigma$ Meson 2000, YITP Kyoto,
June 2000; KEK-Proceedings 2000-4, p. E 115.
\bibitem{REF2} N.N.~Achasov, Nucl. Phys. {\bf A675} (2000), 279c.
\bibitem{REF3} M.D.~Scadron, Proceedings of $\sigma$ Meson 2000, YITP Kyoto,
June 2000; KEK-Proceedings 2000-4, p. E 146.
\bibitem{REF4} N.N.~Achasov, private communication 1/29/02.
\bibitem{REF5} J.D.~Bjorken, private communication 2/07/02.
\bibitem{REF6} E.~Van Beveren {\em et al.}, Modern Phys. Lett. {\bf A17}
(2002), 1673.
\bibitem{REF7} S.N.~Cherry and M.~Pennington, Nucl. Phys. {\bf A688} (2001),
823.
\bibitem{REF8} S.~Narison, private communication 9/19/01; also hep-ph/0208081.
\bibitem{REF9} W.~Ochs and P.~Minkowski, hep-ph/0209223, hep-ph/0209225; also
many private communications.
\bibitem{REF10} M.~B\"{u}scher {\em et al.}, hep-ph/0301126.
\bibitem{REF11} J.L.~Rosner, private communications.
\bibitem{REF12} See S.F.~Tuan, hep-ph/0109191, Hadron Specctroscopy, 
Proceedings of Hadron 2001, p. 495 [D.~Amelin and A.~Zaitsev, Editors A.I.P. (2002).
\bibitem{REF13} K.~Igi, Proceedings of $\sigma$ Meson 2000, YITP Kyoto, June
2000; KEK-Proceedings 2000-4, p. E 79; R.~Kami\'{n}ski {\em et al.}
hep-ph/0210334 also private communication; N.~Isgur and J.~Speth, Phys. Rev.
Lett. {\bf 77} (1996), 2332, and also private communication with J.~Speth.
\bibitem{REF14} V.V.~Anisovich, private communications.
\bibitem{REF15} N.A.~T\"{o}rnqvist, Proceedings of $\sigma$ Meson 2000, YITP
Kyoto, June 2000; KEK-Proceedings 2000-4, p. E 224.
\bibitem{REF16} M.~Ishida, Proceedings of $\sigma$ Meson 2000, YITP Kyoto,
June 2000; KEK-Proceedings 2000-4, p. E 103.
\bibitem{REF17} D.~Ito {\em et al.}, hep-ph/0208245.
\bibitem{REF18} I.~Yamauchi, these Proceedings.
\bibitem{REF19} S.~Stone, private communication 12/04/02.
\bibitem{REF20} V.V.~Anisovich, on hep-ph/0208123, Fig. 11.
\bibitem{REF21} S.L.~Adler, private communication.
\bibitem{REF22} N.A.~T\"{o}rnqvist, hep-ph/0204215 v3 6 Jun 2002.
\bibitem{REF23} T.~Hatsuda and T.~Kunihiro, Prog. Theor. Phys. {\bf 74}
(1985), 765; T.~Kunihiro, Prog. Theor. Phys. {\bf 120} (1995), 75;
T.~Kunihiro, nucl-th/0006035; Y.~Nambu, Physica {\bf D15} (1985) 147 and in
{\em From Symmetries to Strings} (World Sci. '90); {\em top condensation model}
in ``New Theories in Physics'' (World Sci. '89).
\bibitem{REF24} J.D.~Bjorken, hep-th/0111196.
\bibitem{REF25} R.L.~Jaffe and F.E.~Low, Phys. Rev. {\bf D19} (1979), 2105.
\bibitem{REF26} N.I.~Kochelev, private communication 9/22/02.
\bibitem{REF27} M.~Iwasaki, Proceedings of $\sigma$ Meson 2000, YITP Kyoto,
June 2000; KEK-Proceedings 2000-4, p. E 195; S.~Hirenzaki {\em et al.}, Nucl.
Phys. {\bf A710} (2002), 131.
\bibitem{REF28} P.~Jaikumar and I.~Zahed, hep-ph/0209249.
\bibitem{REF29} Dirac Collaboration, $\pi$N News Lett. {\bf 16} (2002) 352.
http://dirac.web.cern.ch/Dirac on Pionium, Kaonium.
\bibitem{REF30} N.N.~Achasov and G.N.~Shestakov, Phys. Rev. {\bf D49} (1994),5779.
\bibitem{REF31} M.~Ishida, hep-ph/0212383 v1 29 Dec 2002, also these Proceedings.
\bibitem{REF32} References given in for instance S.F.~Tuan, hep-ph/0109187, Hadron
Spectroscopy, Proceedings of Hadron 2001, p. 105 [D.~Amelin and A.~Zaitsev,
Editors A.I.P. (2002)].
\bibitem{REF33} M.~Suzuki, private communications.
\bibitem{REF34} CLEO Collaboration, Phys. Rev. Lett. {\bf 78} (1997), 3261.
\bibitem{REF35} J.L.~Rosner and C.W.~Chiang, hep-ph/0212274 v3 8 Jan 2003.
\bibitem{REF36} J.M.~Link {\em et al.} (FOCUS Collaboration), Phys. Lett. {\bf
B535} (2002), 43.
\bibitem{REF37} D.~Aston {\em et al.} (LASS Collaboration), Nucl. Phys. {\bf B296}
(1988), 493.
\bibitem{REF38} E.P.~Shabalin, Yad. Fiz. {\bf 49} (1989), 588 [Sov. J. Nucl. Phys.
{\bf 49} (1989), 365].
\bibitem{REF39} E.M.~Aitala {\em et al.} (E791 Collaboration), Phys. Rev. Lett.
{\bf 16}, 121801; hep-ex/0204018.
\bibitem{REF40} B.T.~Meadows, private communication, 22/10/02.
\bibitem{REF41} CLEO Collaboration, Phys. Rev. Lett. {\bf 89} (2002), 251802.
\bibitem{REF42} BABAR Collaboration, hep-ex/0207089.
\bibitem{REF43} D.V.~Bugg, private communications.
\bibitem{REF44} E.~Van Beveren and G.~Rupp, hep-ph/0201006.
\bibitem{REF45} The work of the Ishida group is given in Phys. Lett. {\bf B518},
31, 47 (2001), hep-ph/0110357, hep-ph/0110358, and hep-ph/0212383.
\bibitem{REF46}T.M.~Yan, Phys. Rev. {\bf D22} (1980), 1652.
\bibitem{REF47} M.~Ishida, private communication, 7/22/02.
\bibitem{REF48} V.V.~Anisovich {\em et al.}, Phys. Rev. {\bf D51} (1995), R4619.
\bibitem{REF49} S.~Stone, private communication, 2/11/03.
\bibitem{REF50} T.~Barnes, hep-ph/0202157, Hadron Spectroscopy, Proceedings of
Hadron 2001, p. 447 [D.~Amelin and A.~Zaitsev, Editors A.I.P. (2002)].
\bibitem{REF51} R.H.~Dalitz {\em et al.}, Phys. Lett. {\bf B213} (1988), 537.
\bibitem{REF52} J.M.~Cornwall and S.F.~Tuan, Phys. Lett. {\bf B136} (1984), 110.
\bibitem{REF53} A.~Zaitsev, these Proceedings.
\bibitem{REF54} S.L.~Olsen, these Proceedings, see also hep-ex/0303006.
\bibitem{REF55} BELLE Collaboration, hep-ex/0302024, see especially Fig.2.
\bibitem{REF56} J.L.~Rosner, hep-ph/0303079.
\bibitem{REF57} Y.~Nambu and G.~Jona-Lasinio, Phys. Rev. {\bf 122} (1961), 345;
Phys, Rev. {\bf 124} (1961) 246.
\bibitem{REF58} S.~Pakvasa and S.F.~Tuan, Phys. Lett. {\bf B459} (1999), 301.
\bibitem{REF59} R.H.~Dalitz, in {\em High Energy Physics} (Gordon and Breach, New
York, 1966), p. 253.
\end{thebibliography}
\end{document}